\documentclass{PoS}
\usepackage{epsfig}
\usepackage{amssymb}
\newcommand{\eq}{Eq.~}
\newcommand{\eqs}{Eqs.~}

\renewcommand{\vec}[1]{{\bf #1}}
\newcommand{\nwc}{\newcommand}

\newcommand{\fig}{Fig.~}
\nwc{\nl}  {\newline}
\nwc{\be}  {\begin{equation}}
\nwc{\ee}  {\end{equation}}
\nwc{\bmu} {\bar{\mu}}
\nwc{\ba}  {\begin{eqnarray}}
\nwc{\ea}  {\end{eqnarray}}
\nwc{\bi}  {\begin{itemize}}
\nwc{\ei}  {\end{itemize}}
\nwc{\nn}  {\nonumber\\}
\nwc{\Tr}  {\mathop{\rm Tr}}
\nwc{\re}  {\mathop{\rm Re}}
\nwc{\im}  {\mathop{\rm Im}}
\nwc{\Hc}  {\mathop{\rm H.c.}}
\nwc{\la}[1]{\label{#1}}
\nwc{\rmi}[1]{{\! \mbox{\scriptsize #1}}}
\nwc{\rmii}[1]{{\mbox{\tiny\rm{#1}}}}
\nwc{\nr}[1]{(\ref{#1})}
\nwc{\fr}[2]{{\frac{#1}{#2}}}
\nwc{\msbar}{\overline{\mbox{\rm MS}}}
\nwc{\lambdamsbar}{\Lambda_{\overline{\rm MS}}}
\newcommand{\Nf}{N_{\rm f}}
\newcommand{\Nc}{N_{\rm c}}
\newcommand{\Tc}{T_{\rm c}}
\newcommand{\se}{Sec.~}

\def\mslash#1{#1\!\!\!\!\!/\!\,\,} 
\def\lsi{\raise0.3ex\hbox{$<$\kern-0.75em\raise-1.1ex\hbox{$\sim$}}}
\def\gsi{\raise0.3ex\hbox{$>$\kern-0.75em\raise-1.1ex\hbox{$\sim$}}}
\nwc{\lsim}{\mathop{\lsi}}
\nwc{\gsim}{\mathop{\gsi}}

\def\NN{{\rm I\kern -.16em N}}
\def\II{{\rm I\kern -.16em I}}
\def\RR{{\rm I\kern -.2em  R}}
\def\ZZ{Z \kern -.43em Z}
\def\QQ{{\rm \kern .25em
             \vrule height1.4ex depth-.12ex width.06em\kern-.31em Q}}
\def\CC{{\rm \kern .25em
             \vrule height1.4ex depth-.12ex width.06em\kern-.31em C}}

\title{Hot QCD and warm dark matter}

\ShortTitle{Hot QCD and warm dark matter}

\author{\speaker{M.~Laine}\\
        Faculty of Physics, University of Bielefeld, 
        D-33501 Bielefeld, Germany \\ 
        E-mail: \email{laine@physik.uni-bielefeld.de}}

\abstract{
One of the possible explanations for the dark matter needed in the
standard cosmological model is so-called warm dark matter, in the form
of right-handed (``sterile'') neutrinos with a mass in the keV range. 
I describe how various properties of QCD at temperatures of a few
hundred MeV play an important role in the theoretical computations
that are needed for consolidating or falsifying this scenario.
In particular the points where lattice QCD could help are underlined.
}

\FullConference{XXIVth International Symposium on Lattice Field Theory\\
                July 23-28, 2006\\
                Tucson, Arizona, USA}

\begin{document}

%
\section{Introduction}

With the on-going heavy ion collision program at RHIC and
the forth-coming one at LHC, a significant amount of resources 
is being devoted to the study of hadronic matter at very 
high energy densities. Yet, it is sometimes difficult to express
the ultimate physics goals that these programs attempt to meet
in a conceptually clean manner. After all, nobody expects that
one would be able to disprove the validity of QCD through 
these programs, while the computational tools that are being
developed need to take into account the rather specific background effects 
and non-equilibrium features that may hamper the experimental
setup, and are thus somewhat specific to these circumstances, 
without necessarily a much wider applicability in other contexts.    

Of course, it might be that the ultimate motivation for the 
heavy ion program does not lie within QCD phenomenology but elsewhere. 
In documents directed towards the general public, 
in particular, it is often argued that the main aim
is to recreate for a moment the conditions
that existed in the very Early Universe. In other words, one is 
hoping to produce a state which is to a good approximation thermalised, 
with a temperature in the range of a few hundred MeV, whereby 
the results should be of interest to the cosmology community. 

It is a fact, though, that most cosmologists 
have shown relatively little interest in 
the physics related to the QCD epoch since the early 1990's.
The main reason is that in order to create cosmological {\em relics}, 
i.e.\ something that could be astronomically observed today 
(say, baryonic matter, dark matter, inhomogeneities in the
various matter distributions), one needs to deviate from 
thermodynamical equilibrium. This does not happen easily in cosmology, 
though, given that the expansion rate is very small compared
with the rates of microscopic processes. In fact, for strongly 
interacting particles, the only possibility of a deviation from
equilibrium is through the existence of a first order phase
transition. But as lattice studies have been indicating since
a long time already (for recent studies, see refs.~\cite{dfp}), 
the transition in physical QCD seems to be 
an analytic crossover rather than a real phase transition
(or, at most, an extremely weak first order transition).
Such a scenario does not appear to lead to any direct
cosmological relics (for a recent review, see ref.~\cite{rev}).

It turns out, however, that this argumentation may be overly pessimistic.
In some sense the situation is analogous to the case that will be
met in the $pp$-experiments at the LHC: the goal is to find 
``New Physics'', be it a Higgs or something more exotic,
but QCD still plays an instrumental role, because it produces
a significant {\em background} that needs to be understood extremely well.

As a first example in this direction, let us mention the 
properties of the primordial gravitational wave background. It is assumed
that inflation generates a background with a certain ``flat'' spectrum.
The spectrum observed today depends, however, also on the history of the 
Universe after the inflationary period; in particular it depends on how
long a certain wavelength spends inside the horizon after 
re-entering it. This in turn is dictated by the thermal history
and the equation-of-state of the matter filling the Universe. 
Consequently, the QCD epoch, during which the expansion rate 
changes by a significant amount, does lead to a distinctive 
feature in the gravitational wave spectrum that could be observed 
today~\cite{djs} (if our instruments were precise enough).

Another example, closer to the topic of this talk, is the 
problem of Dark Matter (DM). Again, the main ``signals'' are
the measurement of the DM relic density and, ultimately, 
the discovery of DM particles. As we will recall in the 
next section, however, the thermodynamics of the QCD epoch
does play an important ``background'' 
role in determining the relic density.

%
\section{The problem of Dark Matter} 

A number of independent observations, 
ranging from the length scales of galaxies (rotation curves), 
through those of galaxy clusters (lensing, large-scale structure), 
up to cosmological scales (anisotropies in the microwave background), 
have been consistently making a case for the existence of non-baryonic 
DM for quite a while already. The amount of DM
is estimated to presently be about 20\% of the total energy density, with
a relative error of at most 5\%~\cite{pdg}. The error gets constantly 
reduced through new observations, and is expected to 
reach the 1\% level through the Planck mission at the latest. 
Of course, all of these observations are indirect, and involve
some uncontrolled systematic errors. Still, the facts that  
very different types of observations produce consistent numbers, 
and that the outcome has remained stable for a long time already, 
make a reasonably credible case for the existence of DM.

It may be amusing to recall the DM relic 
density in absolute units as well. At the current moment
in the Universe expansion, it amounts to 
\be
 e_\rmi{\,DM} \approx 1.1\,\frac{\mbox{GeV}}{\mbox{m}^{3}}
 \;.
\ee
This may appear to be a small number, but is in fact {\em large} 
compared with the current average energy density of the total of all 
known forms of matter, namely 
\be
 e_\rmi{\,Baryon} \approx 0.2\,\frac{\mbox{GeV}}{\mbox{m}^{3}}
 \;. 
\ee

Given that all existing evidence for DM is based on its
gravitational interactions, the more precise nature 
of DM remains unknown.\footnote{%
 In particular, it must be kept in mind the Nature might  
 have chosen an even more exotic explanation for what we conceive as 
 DM than just some novel particles: maybe the problems are on the 
 side of gravity, and it is a modification of the theory of gravitational 
 interactions which explains the features that we associate with DM.
 } 
The most popular candidate is 
so-called Cold Dark Matter (CDM), consisting of particles with a mass
$m \gsim 10$~GeV, related perhaps to supersymmetry or supergravity.
These particles can be called Weakly Interacting Massive Particles, WIMPs. 
The hope would then be to discover CDM not only through astronomical 
observation, but also through directly creating WIMPs 
in future collider experiments like the LHC. 

Unfortunately,  as of today, there is no concrete
evidence in favour of CDM. Therefore the field remains open for 
other candidates as well. In this talk I will concentrate on 
right-handed ``sterile'' neutrinos as candidates for DM. 
Within the see-saw scenario it is commonplace to introduce
massive right-handed neutrinos in any case, 
to explain the experimental fact that at least
two of the active neutrinos have a mass. 
It is   
then only natural to assume that there are in fact three families of 
right-handed neutrinos in total, like there are of the other particles. 
It is the lightest among these, with a mass in the keV range, 
which could possible act as DM~\cite{old,dw}. 
Because of the small mass, and subsequently large average 
energy at any given temperature, 
such dark matter is referred 
to as Warm Dark Matter (WDM) in contrast to CDM.

Now, let us recall briefly
why hot QCD does play a role in the determination of
the DM relic density, 
even though the DM particles themselves are very weakly 
interacting (otherwise, they would not be ``dark''). For CDM, 
for instance, the relic density is determined through the moment
when the WIMPs {\em decouple}, i.e., when their interaction
rate becomes smaller than the rate of Universe expansion. 
The expansion rate, in turn, does depend on the properties of all 
the particles in the plasma, and most of them do feel strong 
interactions. It turns out, in particular, that 
WIMPs of mass $m$ decouple at $T_\rmi{dec} \sim m/25$ 
(see, e.g., ref.~\cite{kt}).
For $m = $ 10...1000~GeV, $T_\rmi{dec} = $ 0.4...40 GeV, 
which indeed is a range where quarks and gluons dominate
the equation-of-state (EOS). Therefore the QCD EOS does affect 
the CDM relic density on a level which is on par with the observational 
accuracy of the future experiments~\cite{olddm2,dm2}.

For right-handed WDM neutrinos, the production mechanism is 
somewhat different from that for CDM: these particles do not simply 
decouple (because their couplings are so weak that they 
never reached equilibrium), 
but their production rate really peaks at a certain temperature. 
The temperature in question depends on the mass of the WDM neutrino, 
but for the keV range mentioned above (see \fig\ref{fig:eplot}
for the observational constraints which make this range to be 
the relevant one), the peak temperature is around 
$T \sim 200$~MeV \cite{dw} (see also \se\ref{se:why}). 
This is just the temperature scale of the QCD phase transition 
or crossover, so that indirect QCD effects on the WDM relic
density are even more dramatic than on the CDM one. 
It is these effects that will concern us in the following.

%
\section{Minimal model for right-handed neutrinos} 

It has perhaps already become apparent that the philosophies 
behind the CDM and WDM scenarios are rather different. In the CDM
case, the hope is that WIMPs exist and offer a window to genuinely
new physics --- the more exotic, the better! 
In the case of right-handed neutrinos, in contrast, 
the philosophy is to be as down-to-earth as possible: 
the Minimal Standard Model (MSM) is completed by adding only 
those degrees of freedom which are necessary in any case
for explaining the established 
experimental facts concerning neutrino oscillations. Otherwise 
the guiding principle, the construction of the most general 
renormalizable Lagrangian invariant under the gauge
symmetry SU(3)$_C \times$SU(2)$_L\times$U(1)$_Y$, 
remains the same as in the MSM. 

Given these principles, the new Lagrangian reads 
\be
 \mathcal{L} = \mathcal{L}_\rmi{MSM} + 
  \fr12 \bar{\tilde N_s} [ i \mslash{\partial} -M_s ] \tilde N_s 
 - [ h_{\alpha s} \bar L_\alpha \tilde \phi\, a_R \tilde N_s + \Hc ]
  \;, \la{nuMSM}
\ee
where the generation indices $s,\alpha = 1,2,3$ are summed over;
$\tilde N_s$ are Majorana fields; 
$M_s$ are real Majorana masses;
$h_{\alpha s}$ are complex Yukawa couplings; 
$L_\alpha$ are lepton doublets; 
and $\tilde \phi \equiv i \tau_2 \phi^*$ is the conjugate Higgs doublet.  
Because of electroweak symmetry breaking, active 
neutrinos have masses in this model. For illustration 
we will work in the corner of the parameter space where the masses 
are given by the see-saw formula as usual,  
$m_{\nu_\alpha} \sim {|h_{\alpha s}|^2 v^2} / {M_s}$,
where $v \simeq 246/\sqrt{2}$~GeV 
$\approx 174$~GeV is the Higgs field vacuum expectation value.

Now, there is nothing particularly new about 
the Lagrangian in \eq\nr{nuMSM}: it is the very
theory that is practically always taken as the starting
point for the description of massive neutrinos. 
Usually, though, the assumption is made that 
the Yukawa couplings $h_{\alpha s}$ are of 
the same order of magnitude as the known ones, 
$|h_{\alpha s}| \lsim 1$. To reproduce neutrino
masses in the range suggested by the oscillation 
experiments via see-saw, leads then to the assumption that 
$M_1$, $M_2$, $M_3 \sim 10^{10}...10^{15}$~GeV. 
Thereby the active neutrino masses are viewed
as a window to GUT-scale physics. 

Esthetically, though, a theory with such parameter
values may look a bit strange: the (renormalized) 
Higgs mass parameter is in any case expected to be 
of the order of the electroweak scale, so one 
may ask why the Majorana masses should behave any differently. 
Clearly, it is possible to
keep the active neutrino masses fixed, if a decrease
in the Majorana masses is accompanied by a decrease
of the neutrino Yukawa couplings in the proportion
dictated by the see-saw formula. In refs.~\cite{abs,as}, 
it was indeed proposed to consider Majorana masses in 
the range $M_1 \sim \mbox{keV}$, $M_2$, $M_3 \sim \mbox{GeV}$, 
and Yukawa couplings in the ranges 
$|h_{\alpha 1}| \lsim 10^{-11}$, 
$|h_{\alpha 2}|, |h_{\alpha 3}| \lsim 10^{-7}$.
Moreover, it was pointed out that this possibility leads
to some phenomenological benefits compared with the usual 
choice, particularly that the right-handed neutrino 
$N_1$ now has a lifetime long enough to serve as a candidate
for DM. The right-handed neutrinos $N_2$, $N_3$
are the ones that induce the observed active neutrino mass
differences through the see-saw formula, and they also
participate in baryogenesis; however, they
decay too fast to serve as DM (in fact, 
they are constrained to decay before the Big Bang 
Nucleosynthesis epoch~\cite{bbn}). Following 
refs.~\cite{abs,as}, we will refer to the extension of 
the MSM with such parameter choices as the ``$\nu$MSM''.

Given that the Yukawa couplings are small and that 
consequently the right-handed neutrinos interact extremely weakly, 
we will refer to them as ``sterile'' neutrinos in the following. 
For future reference, we also define the mixing angles 
$\theta_{\alpha 1}$ as
\be
 \theta_{\alpha 1} \equiv \frac{h_{\alpha 1} v}{M_1}
 \;.
\ee 
The mixing angles to be considered in the following are all small, 
$|\theta_{\alpha 1}| \lsim 10^{-3}$.

%
\section{Primordial production of right-handed neutrinos} 
\la{se:prod}

Even though sterile neutrinos interact very weakly, 
they do get produced in reactions taking place
in the Early Universe. This leads to a non-zero primordial
abundance, as well as a number of potential consequences:

\begin{itemize}

\item[1.] 
The sterile neutrinos are not stable but decay, 
for instance through the channel $N \to \nu\gamma$.
If their life-time is long enough, this results in 
an X-ray signal which could be observed today. 

\item[2.] 
The sterile neutrinos are massive and thus carry
a certain energy density. Provided, again, that their
lifetime is long enough, this could lead to a contribution
to the WDM that plays a role in structure formation.

\item[3.]
Finally, the various CP-violating scatterings of the sterile neutrinos 
might contribute towards the baryon asymmetry that exists today. 

\end{itemize}

It is worth stressing that the conventional case of heavy
Majorana masses, $M_1,M_2,M_3 \sim 10^{10}...10^{15}$~GeV, 
also leads to a certain primordial abundance and, as is 
well-known, to the possibility of generating a baryon asymmetry 
through the mechanism of thermal leptogenesis~\cite{fy}. 
This region of the parameter space does not lead to the first 
two consequences, however, since the heavy neutrinos decay too 
fast to still be around today, as sources of X-rays and DM.

Because of these potential consequences, 
it is important to determine the abundance 
of the sterile neutrinos as a function of their 
mass $M_1$ and mixing angles $\theta_{\alpha 1}$, 
and we now turn to this computation. 
Afterwards, the result of this theoretical computation
can be confronted with the observational constraints
that follow from the absence of any visible X-ray signal and from 
known properties of large-scale structure (cf.\ \se\ref{se:comp}).

The basic mechanism for the creation of WDM neutrinos
is through inelastic scatterings of light particles 
existing in the thermal plasma, followed by an 
active-sterile neutrino transition. These processes were
first analysed in detail in ref.~\cite{dw}, and a number
of refinements can be found in refs.~\cite{dh}--\cite{bo}.
Our analysis is based on refs.~\cite{als,als2}, where the equations
entering the production rate were derived
directly from quantum field theory, allowing for a systematic
investigation of all the effects that play a role, 
to all orders in the strong coupling constant. 

It turns out that, as already mentioned and as demonstrated
later on explicitly, the production rate peaks at temperatures
of the order of $T \sim 200$~MeV $\ll m_W$. Therefore, 
the basic physics that plays a role
can be understood within the Fermi model. 
We start by sketching the ideas with a few graphs, then discuss 
the general form of the equations that appear; however, the precise 
details are left out and can be consulted in refs.~\cite{als,als2}.

An example of a scattering leading to the production 
of a sterile neutrino $N_1$ is shown by the diagram 
\be
\parbox[c]{10cm}{
%


\centerline{
\epsfysize=2.0cm\epsfbox{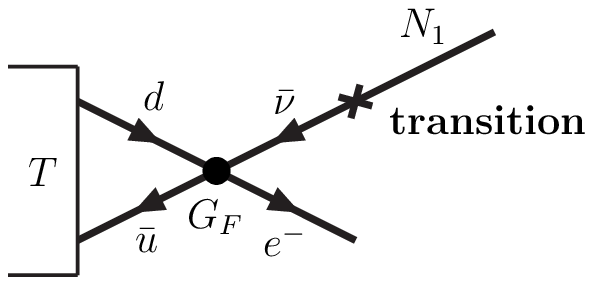}
}


%
}
\;. \la{graphA}
\ee
After their production $N_1$ are essentially inert:
their average lifetime is longer than the age of the Universe, 
and their density is much below the equilibrium value. Therefore, 
the $N_1$ produced essentially escape the thermal system, 
just like photons or dilepton pairs produced 
in heavy ion collision experiments do. 


Now, the rate for the production is proportional to the 
absolute value squared of the amplitude. For the 
process above, this corresponds to the  imaginary part of  
the 2-point function of active neutrinos:
\be
\parbox[c]{10cm}{
%


\centerline{
\epsfysize=2.0cm\epsfbox{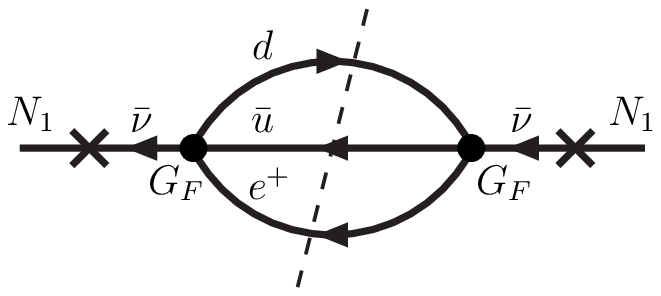} 
}


}
\;, \la{graphB}
\ee
where the dashed line indicates that the cut, or imaginary part, 
is to be taken. 

At temperatures of the order of 200 MeV, however, quarks interact
strongly, and a perturbative evaluation of the 2-loop diagram 
in \eq\nr{graphB} is 
hardly a good approximation. Fortunately, it is possible to 
express its contents in a more general way, whereby the quark
lines can be combined to a propagator corresponding to 
flavour singlet or non-singlet vector or axial current
correlators~\cite{als}: 
\be
\parbox[c]{10cm}{
%


\centerline{
\epsfysize=2.0cm\epsfbox{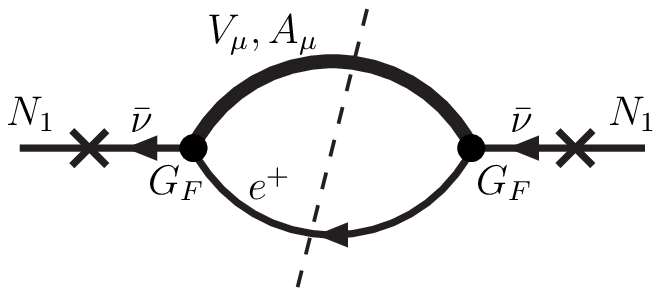}
}


}
\;. \la{graphC}
\ee
The vector and axial current parts of this graph 
can be evaluated also beyond QCD perturbation theory, 
for instance by using chiral perturbation theory 
or lattice techniques (although the latter are faced
with the usual problems related to analytic continuation).

Finally, there would obviously also be a simpler 1-loop
graph that is in principle relevant, namely 
\be
\parbox[c]{10cm}{
%


\centerline{
\epsfysize=1.7cm\epsfbox{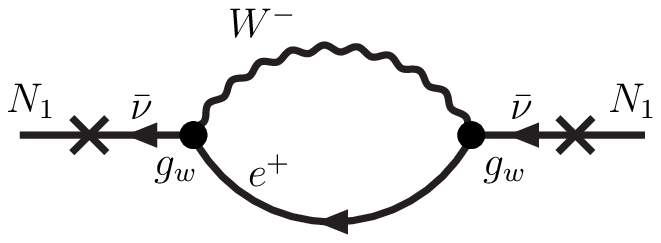}
}


}
\;. \la{graphD}
\ee
The imaginary part of this graph is, however, exponentially
suppressed by $\sim \exp(-m_W/T)$, and thus insignificant at 
the temperatures that we are interested in. On the other hand
this graph does produce a real part (for the real part intermediate 
particles do not need to be on-shell so that no $\exp(-m_W/T)$ appears), 
which indeed plays an important role, as we will see. 

Let us now express these graphs as formulae~\cite{als,als2}. 
Denoting by $n_1$ the number density of the lightest 
sterile neutrinos $N_1$, by $s(T)$ the total entropy density, 
by $e(T)$ the total energy density, and by $c_s^2(T)$ the
speed of sound squared, the production equation can 
be written as 
\be
 -T \frac{{\rm d}}{{\rm d}T}\left[ 
   \frac{n_1(T)}{s(T)}
 \right] 
 \! =  \!
 \frac{1}{3 c_s^2(T) s(T)} 
 \sqrt{\frac{3 m_\rmi{\,Pl}^2}{8\pi e(T)}}
 \int\! {\rm d}^3\vec{q} \, R(T,\vec{q})
 \;, \la{kinetic}
\ee
where $m_\rmi{\,Pl}$ is the Planck mass. The rate, $R(T,\vec{q})$, 
contains in turn the mixing angles and the properties of the 
self-energy $\Sigma_\alpha$ of active neutrinos of flavour $\alpha$: 
\be
 R(T,\vec{q}) 
 \sim 
 \frac{n_\rmi{\,F}(q^0)}{(2\pi)^3 2 |\vec{q}|}
 \sum_{\alpha = e,\mu,\tau} |\theta_{\alpha 1}|^2 
 \frac{M_1^4 \Tr[ \mslash{Q} \im \mslash{\Sigma}_{\alpha} ]}
 {[M_1^2 + 2 |\vec{q}| \re \Sigma_{\alpha}]^2}
 \;. \la{R}
\ee
Here we made use of the fact that sterile neutrinos are on-shell, 
$(q^0)^2 - \vec{q}^2 = M_1^2$.
\eqs\nr{kinetic}, \nr{R} show that we need to determine
$\re\Sigma_{\alpha}$, 
$\im\Sigma_{\alpha}$, as well as the thermodynamic quantities 
$e(T)$, $s(T)$, and $c_s^2(T)$, in order to estimate $n_1(T)$.

Let us start by considering the real and imaginary parts
of the active neutrino self-energy. The real part originates
from the 1-loop graph shown in \eq\nr{graphD};  
a straightforward computation followed by an expansion
in $1/m_W^2$ leads to the result~\cite{ReSigmaold,ReSigma}
\ba
 \re\Sigma_{\alpha} & = & 
 {Q}\, a_{\alpha}(Q) + {u}\, b_{\alpha}(Q)
 \;, \quad u = (1,\vec{0}) \\ 
  b_{\alpha}(Q) & = & \frac{16 G_F^2 T^4}{\pi\alpha_w} q^0
 \left[
   2 \phi\left(\frac{m_{l_\alpha}}{T}\right) 
   + \cos^2\! \theta_\rmii{W}\, 
   \phi\left(\frac{m_{\nu_\alpha}}{T}\right)
 \right]
 \;, \la{bQ}
\ea
where $\phi$ is a simple dimensionless function, which 
can easily be evaluated numerically. The function 
$a_\alpha(Q)$ can in fact be ignored, since this part
of the self-energy is subdominant compared with 
the tree-level term $Q$.

Two interesting remarks can be made related to~\eq\nr{bQ}. 
First of all the parametrically leading term, $\sim 1/m_W^2$, 
vanishes. Therefore the result has a prefactor $G_F^2\sim g_w^4/m_W^4$
(of course only two weak gauge couplings $g_w$ appear, 
whereby we have to divide by $\alpha_w$ after this normalization).
Second, it turns out that if there are non-zero leptonic chemical
potentials in the system, $\mu_L\neq 0$, 
then the leading term, $\sim 1/m_W^2$,
no longer vanishes, but produces a term  
$b_{\alpha}(Q) \sim - G_F n_L$, 
where $n_L$ is the lepton density~\cite{ReSigmaold}. 
This leads to the possibility of a ``pole'', or resonance, 
in \eq\nr{R}, whereby the production rate can be 
enhanced by a significant amount~\cite{ReSigmamu}.
Under normal circumstances, however, it is to be expected
that the lepton density is of the same order of magnitude
as the baryon density, $n_L \sim 10^{-10} T^3$, 
in which case these effects are insignificant
and can be ignored. 
 
As far as the imaginary part of the active neutrino self-energy
in \eq\nr{graphC} is concerned, its general structure is 
\ba
 \im\mslash{\Sigma}_{\alpha}(Q) \!\! & \sim & \!\!  
 G_F^2 \int\! \frac{{\rm d}^3 \vec{r}}{(2\pi)^3}
 \; \mathcal{K}\biggl( \frac{|\vec{q}|}{T}, \frac{|\vec{q}+\vec{r}|}{T}\biggr)
 \gamma^\mu (\mslash{Q} + \mslash{R}) \gamma^\nu 
 \, 
 \rho_{\mu\nu}^{V,A}(|\vec{q} + \vec{r}| - |\vec{q}|,\vec{r})
 \;.
\ea
Here $\vec{r}$ is the spatial momentum that flows through
the vector or axial current propagator, and $\mathcal{K}$ is a 
known thermal ``kernel'', consisting of hyperbolic functions
(cf.\ ref.~\cite{als}). The functions $\rho^{V,A}_{\mu\nu}$ 
represent flavour-singlet and non-singlet vector and axial 
current spectral functions (i.e.,\ imaginary parts
of the retarded two-point functions, $\rho = \im \Pi_R$).

Now, the spectral functions contain both leptonic and hadronic
contributions. In principle these can be computed perturbatively; 
this leads to structures familiar from the source terms of 
Boltzmann equations. For instance, for the case depicted 
in \eq\nr{graphB},
%
\begin{figure}[h]


\centerline{
\epsfysize=8.0cm\epsfbox{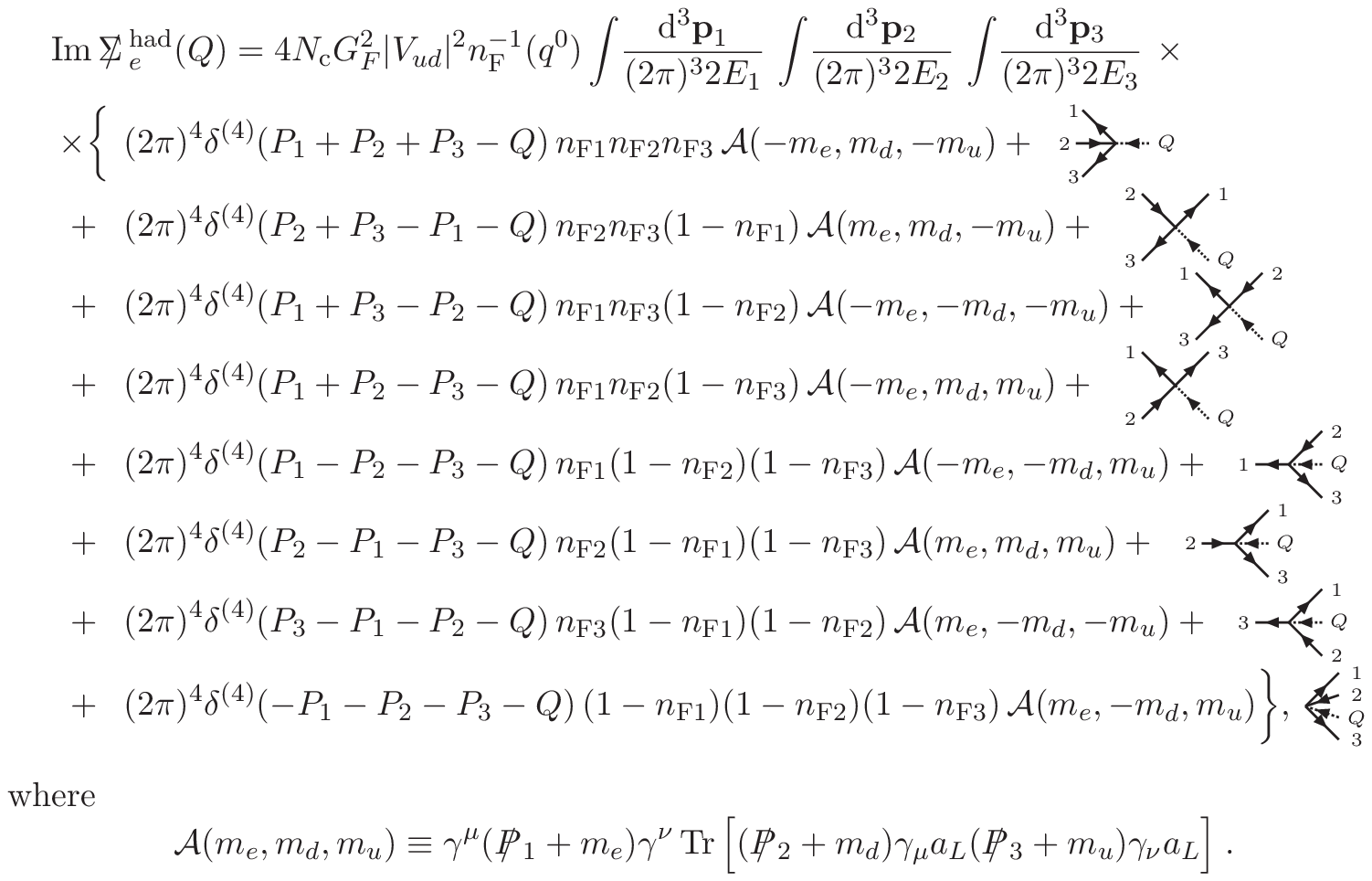}
}


\end{figure}
%

\noindent
Here the labels mean 
$1\equiv e$, 
$2\equiv d$, 
$3\equiv u$.
The hadronic contributions should, however, really
be evaluated beyond perturbation theory, as we have already stressed.

%
\section{Why does the production rate peak at $T \sim 200$~MeV?}
\la{se:why}

Let us recall at this point why the production rate of \eq\nr{R} 
peaks at temperatures of the order of a few hundred MeV~\cite{dw}.

We have already seen that the real part of 
the active neutrino self-energy is of the order 
$
 \re\Sigma_{\alpha} \sim G_F^2 T^4 \alpha_w^{-1} |\vec{q}| 
$
(note that $q^0 \sim |\vec{q}|$ for $|\vec{q}| \gg M_1$ as will be 
the case here). The imaginary part is more complicated, but ignoring all
masses compared with the temperature, it has dimensionally
the form 
$
 \Tr[ \mslash{Q} \im \mslash{\Sigma}_{\alpha}] \sim
 n_\rmi{F}^{-1}(q^0) G_F^2 T^6 f(|\vec{q}|/T)
$,
where $f$ is a non-trivial dimensionless function, numerically
of order unity. It turns out that the rate peaks at momenta of 
order $|\vec{q}| \sim T$, whereby we can replace $f$ by a number
of order unity. Thereby the rate becomes
\be
  R(T,\vec{q}) \propto
 \sum_{\alpha = e,\mu,\tau} |\theta_{\alpha 1}|^2 
 \frac{M_1^4 G_F^2 T^6}{(M_1^2 + 100 G_F^2 T^6)^2}
 \;.
\ee
It is immediately seen that the result is strongly peaked around
temperatures where the two terms in the denominator are of 
similar orders of magnitude, i.e.\  
\be
 T \sim \left( \frac{M_1}{10\, G_F} \right)^{\fr13} 
 \sim 200 \; \mbox{MeV} 
 \left( \frac{M_1}{1\,\mbox{keV}} \right)^{\fr13}
 \;.
\ee
Thus, for sterile neutrino masses in the keV range, the 
production rate accidentally coincides with more or less the QCD scale. 

%
\section{The role of the QCD equation-of-state}
\la{se:eos}

Having analysed $\re\Sigma_{\alpha}$ and $\im\Sigma_{\alpha}$, 
we still need to discuss the thermodynamic functions $c_s^2(T)$, 
$s(T)$ and $e(T)$, in order to be able to integrate \eq\nr{kinetic}.

We parameterise the energy density $e(T)$ 
and the entropy density $s(T)$ through 
effective numbers of bosonic degrees of freedom, 
\ba
 {e(T)}  \equiv  g_\rmi{eff}(T) \frac{\pi^2 T^4}{30}
 \;, \quad
 {s(T)}  \equiv  h_\rmi{eff}(T) \frac{2\pi^2 T^3}{45}
 \;. 
 \la{ieff} 
\ea
The energy and entropy densities follow both from the 
thermodynamic pressure $p(T)$ through standard relations, 
$e(T) = Tp'(T) - p(T)$ and $s(T) = p'(T)$. 
Furthermore the speed of sound squared can be written as
\ba
 c_s^2(T) & \equiv & \frac{p'(T)}{e'(T)} = \frac{p'(T)}{Tp''(T)}
 \;. \la{csT}
\ea
In the non-interacting limit, $c_s^2(T)$ equals 1/3. To summarise, 
we need reliable estimates of $p(T)$, $p'(T)$ and $p''(T)$ in order
to determine the sterile neutrino abundance from \eq\nr{kinetic}.

A reliable determination of $p(T)$ is a long-standing challenge
for finite-temperature field theory. Again, leptonic contributions
can be well treated in perturbation theory, while hadronic contributions, 
which dominate the structure in $p(T)$ in the temperature range of 
interest, are in general hard to compute precisely. 

Certain limiting values of $p(T)$ are understood better, though. 
At low temperatures, \linebreak $T \lsim 100$~MeV,  confinement and 
chiral symmetry  breaking guarantee that the system is composed
of weakly interacting massive hadrons. Treating them as a ``gas''
of resonances, one can at very low temperatures approximate
\be
 p(T) \approx \sum_i T^4 \left( \frac{m_i}{2\pi T} \right)^{\fr32}
 e^{-\frac{m_i}{T}}
 \;,
\ee 
and at somewhat higher temperatures replace this with the 
corresponding relativistic formulae for bosons and fermions, 
respectively. Even though this prescription is rather
phenomenological, lattice simulations suggest that 
it may work surprisingly well
even up to $T\sim 200$~MeV~\cite{krt}.

For high temperatures on the other hand, 
$T \gsim 1000 \mbox{~MeV}$, asymptotic freedom guarantees
that the system can be viewed as a collection of 
weakly interacting quarks and gluons: 
\be
  p(T) \approx \frac{\pi^2 T^4}{90} \left[ 2 ( 
 \Nc^2-1) + \fr72 \Nc \Nf
 \right]\left( 1 + ... + O(g^6) \right)
 \;,
\ee
where $\Nc = 3$ is the number of colours and 
$\Nf$ is the number of massless flavours that play a role
at the temperatures we are interested in. Perturbative
corrections are known up to 4-loop order, 
${O}(g^6)$~\cite{gsixg,nspt}, apart from 
a single missing coefficient. Moreover, quark mass
effects, important for phenomenological applications, 
can be incorporated~\cite{pheneos}.

In between these two limits the situation is much more complicated. 
Ideally, one would like to make use of lattice simulations of the type
in Refs.~\cite{Nfcp}--\cite{Nfbi}.
Unfortunately, it appears that the present results are reliable
in a fairly narrow temperature range only; for instance, the 
characteristic peak that can be seen in $c_s^2$ at around
$T \sim 70$~MeV, due to light pions
(cf.\ \fig\ref{fig:qcdeos}(right)), is not visible
in the existing simulations. In fact  the simulations
display rather a somewhat deeper dip (down to $\sim 0.1$)
around the critical region, and then rise at most slightly as the 
temperature is lowered. 

For these reasons, the results that will be presented in 
the following adhere to the procedure introduced in ref.~\cite{pheneos}, 
rather than to lattice simulations. This procedure makes 
use of a gas of hadronic resonances at low temperatures; 
the most advanced (up to resummed 4-loop level~\cite{gsixg}) weak-coupling 
results at high temperatures; and an interpolation thereof at intermediate
temperatures.\footnote{%
  We have corrected a minor error in 
  the numerical results of ref.~\cite{pheneos}.} 
Remarkably, the temperature interval where an interpolating
function is needed in order to sew together the two asymptotic functions
is fairly narrow, not more than $(10...20)$~MeV, and centered around 
$T \approx 200$~MeV, in curiously good agreement with the crossover
temperature $\Tc \simeq 192 \pm 8$~MeV as suggested by recent
large scale lattice simulations~\cite{Tc}.

\begin{figure}[t]

\centerline{%
~~\epsfysize=5.0cm\epsfbox{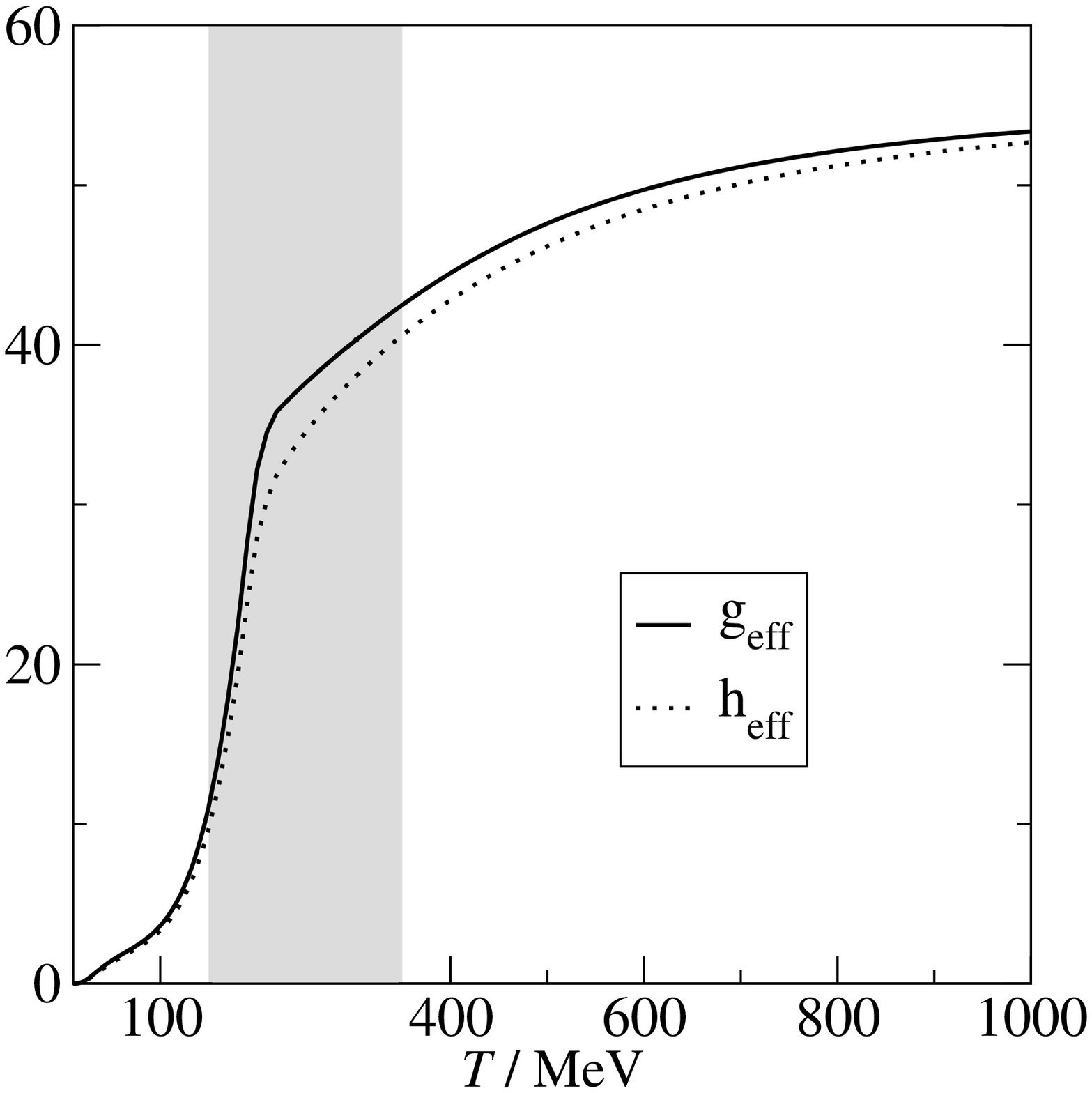}%
~~\epsfysize=5.0cm\epsfbox{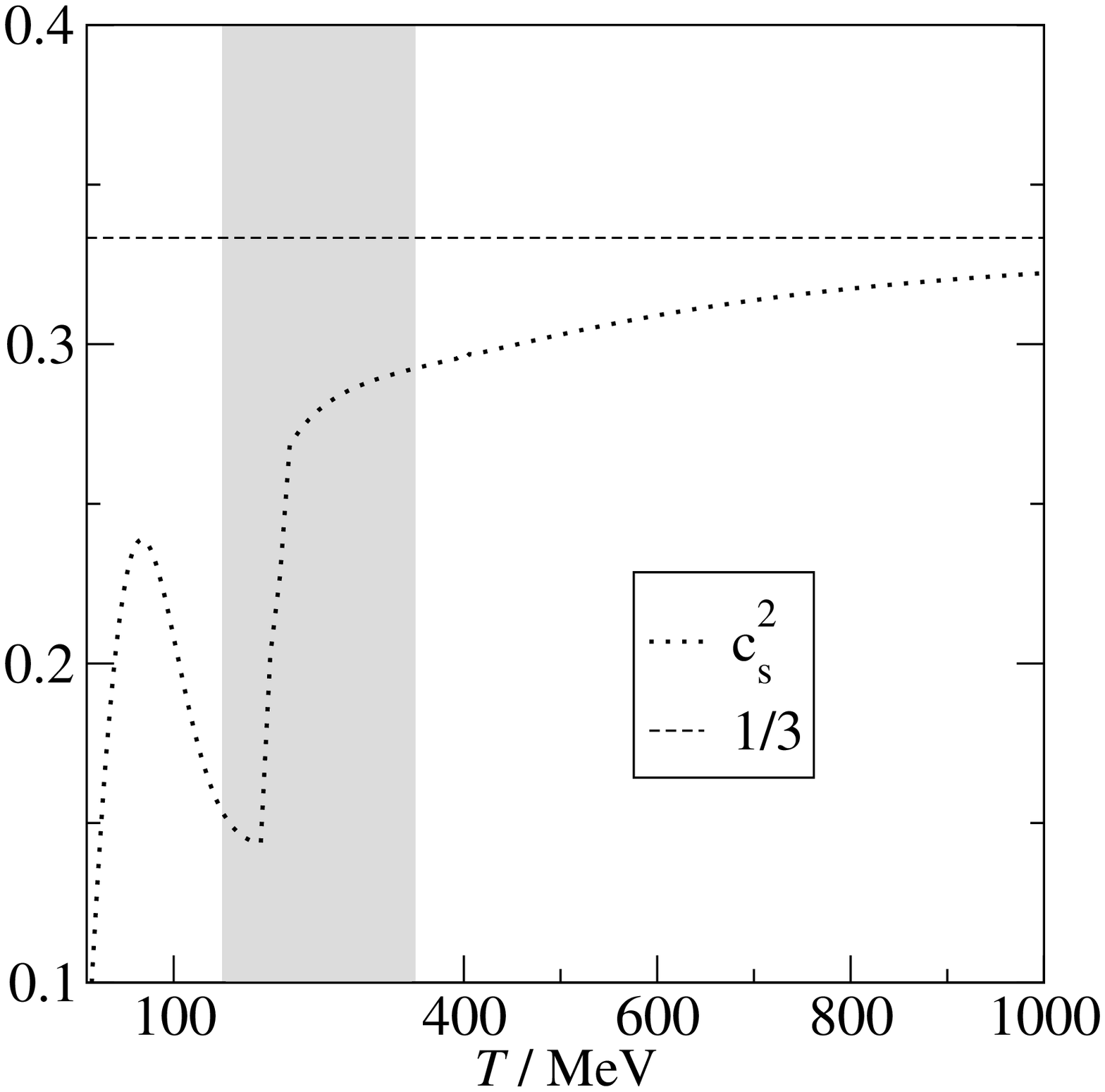}%
}

\caption[a]{
Left:  $g_\rmi{eff}, h_\rmi{eff}$ as defined 
in~\eq\nr{ieff}, for $\Nf = 4$ QCD with physical quark masses~\cite{pheneos}.
Right: the speed of sound squared $c_s^2$, for the same system. 
The shaded region is the range of temperatures where our
recipe is purely phenomenological and needs to be improved
through future lattice simulations.
} 

\la{fig:qcdeos}
\la{fig:qcdpoT4}
\end{figure}

\begin{figure}[t]

\centerline{%
~~\epsfysize=5.0cm\epsfbox{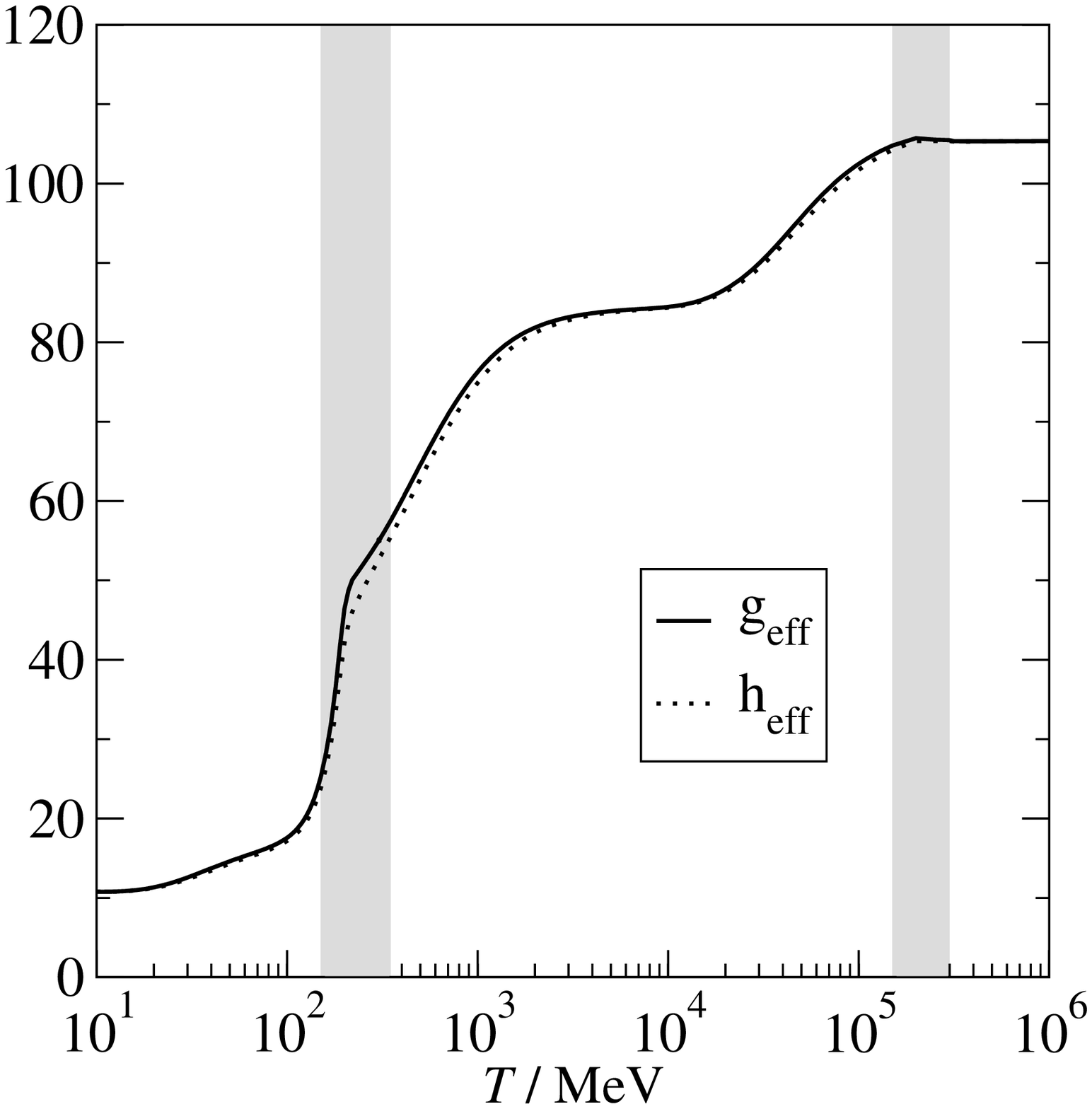}%
~~\epsfysize=5.0cm\epsfbox{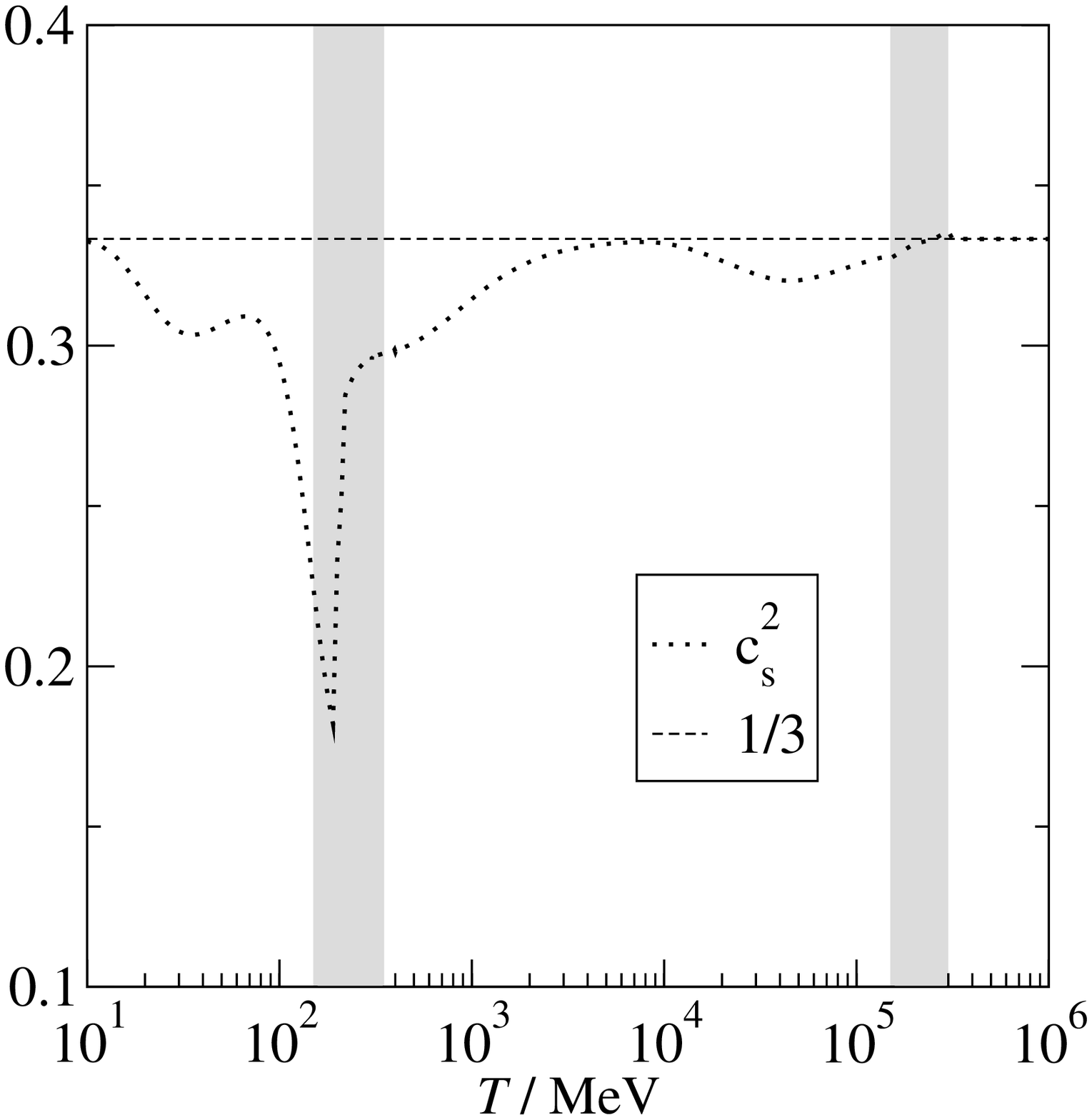}%
}

\caption[a]{
Left:  $g_\rmi{eff}, h_\rmi{eff}$ as defined 
in~\eq\nr{ieff}, for the Standard Model with $m_H =$ 150~GeV~\cite{pheneos}.
Right: the speed of sound squared $c_s^2$, for the same system.
The shaded regions are the ranges of temperatures where our
recipe is purely phenomenological and needs to be improved
through future lattice simulations.
} 

\la{fig:eweos}
\la{fig:ewpoT4}
\end{figure}

The results that follow from this recipe for the quantities 
that are important for us are illustrated in \fig\ref{fig:qcdeos} 
for the QCD part, and in \fig\ref{fig:eweos} for the whole Standard Model.
In the latter case we have displayed, for completeness, the results
in a very broad temperature range.

%
\section{Numerical results} 

With the ingredients discussed, 
\eq\nr{kinetic} can be solved numerically, as a function of 
the temperature. From the result, $n_1(T)/s(T)$, we can 
derive the current energy density, $M_1 n_1(T)$, relative
to the current entropy density. Finally, the current
entropy density, whose value is well-known~\cite{pdg}, can be
traded for the current critical energy density. 
Thereby we obtain the parameter $\Omega_{N_1}$, 
characterising the fraction of the current energy
density that is carried by sterile neutrinos. 
This result can then be compared with the observed 
value for $\Omega_\rmi{\,DM}$.  

In \fig\ref{fig:ex} we show an example of a solution, 
for specific parameter values~\cite{als2}. We have here considered the 
contribution from the active flavour $\alpha = e$ only. 
It can be observed that the sterile neutrinos indeed
get generated at temperatures of a few hundred MeV, 
and that their relic density can be of an order of magnitude 
which is relevant for the explanation of DM.

On the other hand, as can be seen in \fig\ref{fig:ex},
the current results contain relatively significant hadronic
uncertainties. These originate primarily from two sources.  
First of all, as discussed in Section~\ref{se:prod}, 
the imaginary part of the active neutrino self-energy
contains the spectral functions of vector and axial
currents, which involve poorly understood hadronic 
effects. It is of course clear that at low enough temperatures
there are no hadronic scatterings taking place, since
all hadrons are massive. At the same time, at high enough
temperatures, the hadronic contributions can be computed
perturbatively. A very conservative way to estimate 
the ``error'' is to consider these as limiting cases: as a lower bound
no hadrons ($\Nc = 0$), as an upper bound (almost) free quarks.
In the future
this error can hopefully be reduced with the help of lattice 
simulations; the current status on estimating some of the relevant 
spectral functions from the lattice is reviewed in ref.~\cite{th}.

\begin{figure}[tb]


\centerline{\epsfysize=7.0cm\epsfbox{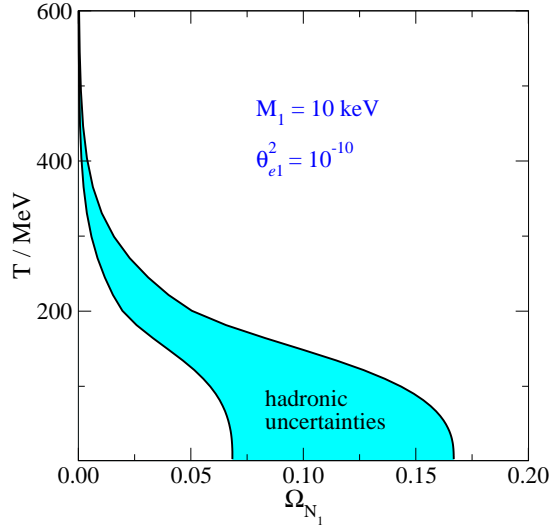}}


\caption[a]{An example of a numerical solution 
of \eq\nr{kinetic} (according to ref.~\cite{als2}), 
normalised with respect to the present total critical energy 
density $e_\rmi{cr}$: 
$
 \Omega_{N_1}(T) \equiv M_1 n_1(T)/s(T) \times 
 s(T_\rmi{today})/ e_\rmi{cr}(T_\rmi{today})
$.}

\la{fig:ex}
\end{figure}

Second, as discussed in Section~\ref{se:eos}, 
the hadronic equation-of-state contains significant
uncertainties in the temperature range of interest. 
It turns out that the most important effect as far
as the current computation is concerned is the location
of the (pseudo)critical temperature $\Tc$: whether the 
production rate peaks above or below this temperature has a large 
effect on the final result, given that the kinetic 
equation of \eq\nr{kinetic} is inversely proportional
to $h_\rmi{eff}\, g_\rmi{eff}^{1/2}$. We have estimated
these uncertainties by rescaling the temperature units
by 20\% in either direction, which certainly is a conservative
estimate. The current status of lattice determinations
of various thermodynamic quantities in the vicinity of $\Tc$
is summarised in ref.~\cite{uh}.

The band indicated in \fig\ref{fig:ex} 
incorporates both of the error sources discussed, 
and provides for a conservative estimate  of the 
possible hadronic uncertainties in the results. 

%
\section{Comparison with observational constraints} 
\la{se:comp}

We can now confront the theoretical result of 
\fig\ref{fig:ex} with observational constraints. 
There are observational constraints from two sides. 
First of all, if sterile neutrinos with a very small
mass constitute all of DM, then structures
on small scales tend to be wiped out, compared with 
structure formation simulations carried out with CDM. 
Comparing the outcome with actual data, particularly
in the form of so-called Lyman-$\alpha$ forest observations, 
which are sensitive to the smallest distance scales, puts 
thus a lower bound on the mass of the sterile 
neutrinos~\cite{hlps}--\cite{mr2}. 
It appears that the lower bound could be as  
high as (8...12)~keV~\cite{als2}. 

Second, the heavier the sterile neutrinos are, 
and the bigger their mixing angles $|\theta_{\alpha 1}|$
with active neutrinos, the more likely are they to decay. 
This leads to a characteristic X-ray signal, which however
has not been observed. Therefore, it is possible to set
an upper bound on a combination of the mass and the mixing
angles of the sterile neutrinos~\cite{dh,aft,bnr}.

\begin{figure}[t]


\centerline{\epsfysize=7.0cm\epsfbox{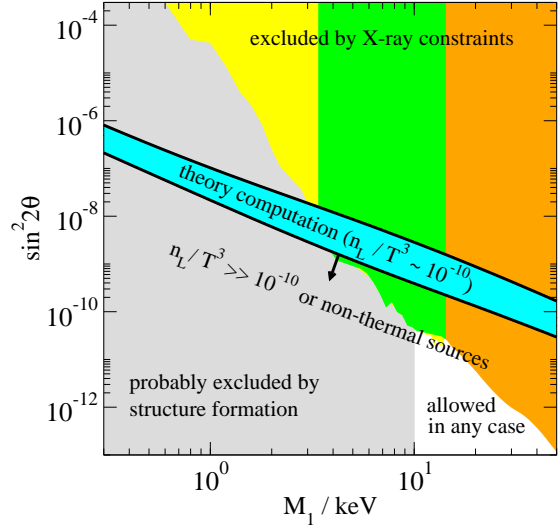}}


\caption[a]{An exclusion plot for the scenario where 
sterile neutrinos constitute all of dark matter 
(adapted from ref.~\cite{als2}; see there also for 
more details). We compare observational constraints 
with the theoretical computation, with the size of hadronic uncertainties
indicated with the band. Here 
$\theta\equiv\sqrt{\sum_{\alpha = e,\mu,\tau}|\theta_{\alpha 1}|^2}$. 
}

\la{fig:eplot}
\end{figure}

These two sets of observational constraints are illustrated 
in \fig\ref{fig:eplot}, together with the result
of the theoretical computation outlined above. 
The basic feature to be observed is that 
the theoretical computation appears to require 
a {\em larger} mixing angle than experimentally allowed, 
in order to produce the observed amount of DM
(for $n_L/T^3 \sim 10^{-10}$). It should be mentioned, however, 
that the structure formation constraints include a number of 
systematic uncertainties, both on the side of the observational
Lyman-$\alpha$ data, and for the theoretical reason that 
they assumed a thermal shape for the momentum distribution function
of sterile neutrinos, even though these are out-of-equilibrium.
If the bound happens to be correct despite these uncertainties, 
\fig\ref{fig:eplot} indicates that sterile neutrinos can only
act as dark matter if there is a lepton asymmetry
$n_L/T^3 \gg 10^{-10}$ in the Universe, or if there are 
additional production mechanisms
apart from  thermal scatterings; it could be, for instance, that a certain 
number density was produced during the inflationary period already~\cite{st}, 
to which thermal scatterings would then {\em add} their contribution.
In any case, the band around the theoretical curve in \fig\ref{fig:eplot} 
indicates the important role that hadronic effects can play.

%
\section{Conclusions}

The basic scenario of right-handed neutrinos serving
as warm dark matter is an old one by now~\cite{old,dw}. 
It has experienced quite a revival recently, though, 
because of significant progress both on the observational
and on the theoretical sides. 

On the observational side, the parameter space
(mass, mixing angles) for the lightest right-handed neutrino
has been strongly constrained by structure formation
and X-ray bounds during the last year or so. This means
that attention can now be focussed on a rather specific situation.

On the theoretical side, the scenario of right-handed neutrinos
serving as dark matter has taken a more prominent role, 
thanks to the realization that there is a minimal model, 
the ``$\nu$MSM'', which not only addresses the dark matter 
problem, but may also explain neutrino
masses~\cite{abs}, baryogenesis~\cite{ars,as}, 
and perhaps also various astrophysical problems~\cite{astro}.
Furthermore, the theoretical tools that are needed in the dark matter 
computation have reached a more mature level~\cite{als}.

These two sides imply that theoretical computations need
to be promoted to a higher level of accuracy than before. 
As has been underlined in this talk, a fully satisfactory 
analysis in this respect is only possible once lattice
studies of the QCD equation-of-state and of vector and axial
current spectral functions with various flavour structures, 
produce results with controlled statistical and systematic 
errors, such that the band in \fig\ref{fig:eplot} can be made 
narrower.  Currently the hadronic uncertainties are of order 50\%
(cf.\ \fig\ref{fig:ex}),  and it would certainly
be desirable to at least half this uncertainty.

%
\section*{Acknowledgments}

This talk is based on refs.~\cite{als,als2,pheneos}; 
I wish to thank Takehiko Asaka, Misha Shaposhnikov and 
York Schr\"oder for collaboration.

%

\end{document}